\documentclass[a4,DIV=16]{scrartcl}

\usepackage{mathrsfs}
\usepackage{tabularx}
\usepackage{graphicx}
\usepackage{amsmath,amssymb,amsfonts}

\def\RR{{\mathbb R}}

\def\be{\begin{equation}}
\def\beq#1{\begin{equation}\label{#1}}
\def\ee{\end{equation}}
\def\bea{\begin{eqnarray}}
\def\beqa#1{\begin{eqnarray}\label{#1}}
\def\eea{\end{eqnarray}}
\def\ba{\begin{array}}
\def\ea{\end{array}}

\def\cL{{\mathcal L}}

\def\cN{{\mathcal N}}

\def\bds{_{-\infty}^\infty}

\def\bbf{{\boldsymbol f}}

\def\bx{{\boldsymbol x}}
\def\by{{\boldsymbol y}}

\def\lab{^k_{gc}}

\newtheorem{remark}{Remark}

\renewcommand\cL{{\mathscr L}}

\newcommand{\sbs}[1]{\ensuremath{_{\textrm{#1}}}}
\def\dsm{\displaystyle}

\begin{document}

\title{A Bayesian model for microarray datasets merging}
\author{Marie-Christine Roubaud and Bruno Torr\'esani$^*$\\{}\\ {\small Aix-Marseille Universit\'e, CNRS, Centrale Marseille, LATP, UMR 7353, 13453 Marseille, France}}
\date{April 2012}

\maketitle

\begin{abstract}

The aggregation of microarray datasets originating from different studies is
still a difficult open problem. Currently, best results are generally obtained
by the so-called meta-analysis approach, which aggregates results from
individual datasets, instead of analyzing aggregated datasets. In order to
tackle such aggregation problems, it is necessary to correct for interstudy
variability prior to aggregation. The goal of this paper is to present a new
approach for microarray datasets merging, based upon explicit modeling of
interstudy variability and gene variability.

We develop and demonstrate a new algorithm for microarray datasets merging. The
underlying model assumes normally distributed {\em intrinsic gene expressions},
distorted by a study-dependent nonlinear transformation, and study dependent
(normally distributed) observation noise. The algorithm addresses both
parameter estimation (the parameters being gene expression means and variances,
observation noise variances and the nonlinear transformations) and data
adjustment, and yields as a result
adjusted datasets suitable for aggregation.

The method is validated on two case studies. The first one concerns {\em E.
  Coli} expression data, artificially distorted by given nonlinear
transformations and additive observation noise. The proposed method is able to
correct for the distortion, and yields adjusted datasets from which the relevant
biological effects can be recovered, as shown by a standard differential
analysis.
The second case study concerns the aggregation of two real prostate cancer
datasets. After adjustment using the proposed algorithm, a differential analysis
performed on adjusted datasets yields a larger number of differentially
expressed genes (between control and tumor data). 

The proposed method has been implemented using the statistical software {\tt R}
\footnote{{\tt www.r-project.org}, the R Project for Statistical Computing.},
and {\sc Bioconductor} packages~\footnote{{\tt www.bioconductor.org},
  Bioconductor, opensource software for bioinformatics.}.
The source code (valid for merging two datasets), as well as the datasets used
for the validation, and some complementary results, are made available on the
web site 

\centerline{\footnotesize \tt www.latp.univ-mrs.fr/$\sim$mcroubau/MicroarrayMerging}

\end{abstract}

\section{Introduction}
Microarray technology provides high throughput measurements of messenger
RNA levels of thousands of genes in tissue samples. Since its introduction almost
twenty years ago, it has found applications in many aspects of molecular
genetics and functional genomics, from the discovery of basic biological
mechanisms to the classification of diseases into subgroups and the prediction
of disease outcome. 

However, although important progress has been made, that has turned lab
experiments into industrially standardized protocols, the technology is still
facing significant reproducibility problems (see e.g. the introduction
of~\cite{Ramasamy09key} for a more detailed account). Therefore, comparing
results from different studies is generally difficult, and extreme care is
needed if one wishes to merge different datasets into a single one to improve
reliability and generalizability of the results.

Several microarray study aggregation techniques have been proposed. Among them,
two approaches have received significant attention. The first one (see for
example~\cite{Hong08comparison}) proposes to
replace numerical expression data with ranked data, in order to correct for
study-dependent normalization effects. This implicitely assumes that the ranks
are essentially respected in the considered datasets; unfortunately, such an
assumption fails to be true in many situations, for example when a large number
of weakly expressed genes are present, the fluctuations of which induce large
deviations in ranks. The second classical approach, termed {\em Meta-Analysis},
aggregates the results of statistical analyzes from the different datasets
rather than aggregating the datasets themselves (see for
example~\cite{Cahan07metaanalysis} for a review). It has been argued by various
authors (see for example~\cite{Cahan07metaanalysis,Conlon07bayesian}) that such
meta analysis approaches generally outperform classical statistical analysis
directly performed on aggregated datasets.

We propose in this paper a new approach for pre-processing gene expression
datasets prior to aggregation, as an alternative to meta-analysis based
techniques. The idea is to start with an explicit model for gene expression
data, and an explicit model for the distortions induced by the different
experiments to be merged. In a Bayesian framework, those two models are used as
priors. Assuming normally distributed additive observation noises leads to
a closed form expression for the posterior probability.

This model is simple enough to yield a
fairly standard estimation algorithm, exploiting elementary linearization
and optimization techniques. The latter yields estimates for the model 
parameters, together with adjusted gene expression values in which both the
non-linear distortions and the variances inhomogenities have been corrected,
therefore suitable for aggregation and further statistical analysis.

The model and the algorithm are validated in both synthetic datasets (i.e. real
datasets with an artificial distortion and additional noise), for which the
method is shown to be able to  correct for the distortion and noise (to some
extent). We then apply the approach to real prostate cancer expression datasets,
for which two original datasets are aggregated after correction.

Preliminary results of this work have been given in conference
proceedings~\cite{Roubaud11new}.

\section{Approach}
\label{se:approach}
The approach we develop here is based upon an explicit modeling of the gene
expressions, and of the distortion introduced by the several studies. For the
sake of simplicity we limit ourselves to a simple model for logarithms of gene
expressions, assuming independent normally distributed values (with unknown
means and variances). We also assume that each study gives rise to a non-linear
distortion (called an observation function) of gene expressions, and an additive
zero-mean Gaussian white noise, of unknown variance. Our approach mainly relies
on two steps. First, estimate the parameters of the model (including the nonlinear
observation functions), then estimate adjusted gene expression values, that can
be aggregated into a single dataset.

\medskip
More precisely, we rely on the following model for logarithms of measured gene
expression data. Assume we are given several datasets $k=1,\dots K$, involving
$g=1,\dots N_g$ genes (if necessary, after restriction to a suitable subset of
common available genes). 
For each dataset $k$, several arrays (hereafter termed {\em conditions}
$c=1,\dots N^k_c$ are available. We denote by $y\lab$ the logarithms of measured
gene expression levels, which we shall call throughout this paper the {\em
  observed gene expressions}. The ingredients of the model are the following:
\begin{itemize}
\item
The {\em intrinsic gene expression} values $\bx = \{x\lab\}$ are assumed to be
independent realizations of i.i.d. normally distributed random variables
$X_g\sim\cN(\mu_g,\sigma_g^2)$ (whose distributions do not depend on the study
$k$). We write
$$
x^k_{gc} = \mu_g + \delta\lab\ .
$$
\item
The {\em observed gene expression} values $\by = \{y\lab\}$ are obtained by a
study-dependent nonlinear transformation $f_k$ of the intrinsic expression values,
with additive Gaussian noise, $Y^k_g = f_k(X_g) + U^k$, with
$U^k\sim\cN(0,\tau_{k}^2)$. Componentwise, the observations therefore read
$$
y^k_{gc} = f_k(x^k_{gc}) + u^k_{gc}\ .
$$
\item
The {\em observation functions :} $\bbf = \{f_k,\ k=1,\dots K\}$ are assumed to
be monotonic, smooth functions ($\int |f_k''(x)|^2\,dx<\infty$), and are given a
spline-type prior distribution 
$$
p(f_k)\sim \exp\left\{-\lambda_k\int\bds |f_k''(x)|^2dx\right\}\ .
$$
\end{itemize}
Conditionally to the observation functions $f_k$ and the intrinsic gene
expressions $x\lab$, the observed gene expressions $y\lab$ are
therefore independent and normally distributed, with means $f_k(x\lab)$ and
variances $\tau_k^2$. 
Applying the Bayes rule, it is easily seen that the log posterior probability
distribution is (up to an additive constant) equal to the sum of three components
$$
\cL(\bx,\bbf | \by) =\cL^{(1)} + \cL^{(2)} + \cL^{(3)}\ ,\qquad
\hbox{with}
$$
\begin{equation}
\left\{
\begin{array}{lll}
\cL^{(1)}\! &\!\!=\! \dsm{\sum_{k,g} \cL^{(1);k}_g}\!\!
&\!\! =\! - \dsm{\sum_{k,g}\frac1{2\tau_k^2}\sum_c[y\lab - f_k(x\lab)]^2}\\
\cL^{(2)} &\!\! =\!  {\dsm\sum_{k,g} \cL^{(2);k}_g }\!\!
&\!\!=\! -\dsm{\sum_g\frac1{2\sigma_g^2}\sum_{k,c}[x\lab - \mu_g]^2}\\
\cL^{(3)} &\! =\!\! {\dsm \sum_{k} \cL^{(3);k} }\!\!
&\!\!=\!-\dsm{\sum_k\lambda_k\int\bds |f_k''(x)|^2\,dx}
\end{array}
\right.
\label{eq:logposterior}
\end{equation}
The approach we develop here aims at maximizing numerically the above
log-posterior probability, estimating the gene average expressions
$\mu_g$ and variances $\sigma_g^2$, the observation noise variances $\tau_k^2$
and the observation functions $f_k$, and solve the regression problem for
estimating intrinsic gene expression values $x\lab$.

\section{Methods}
We describe in this section the estimation method used in our approach, together
with the corresponding algorithm. The first step of the algorithm is the
estimation of the model parameters: means, variances and observation function.
The chosen approach is an iterative one, each parameter is estimated at the
time.
After the estimation, adjusted gene expression datasets are available, that can
be aggregated and analyzed.

\subsection{Estimation}
\subsubsection{Observation functions estimation}
\label{sub:obsfuncest}
Suppose firstly that the {\em intrinsic} gene expression values $x\lab$ are
fixed. Then the observation function estimation problem
reduces to the minimization with respect to
$\bbf=\{f_k\in H^2(\RR),k=1,\dots K\}$ of the quantity
$$
\Gamma[\bbf]= \sum_{k}\left[\left(\sum_g \cL^{(1);k}_g\right)  + \cL^{(3);k} \right]
$$
and decouples further into the following $K$ independent optimization problems:
for $k=1,\dots K$, solve
$$
f_k = {\rm arg}\min_{f} \left\{\frac1{\tau_k^2} \sum_{g} \left[y\lab - f(x\lab)\right]^2
+ \lambda_k\int |f''(x)|^2\,dx\right\}\ .
$$

The latter are actually smoothing spline regression problems (see
e.g.~\cite{Wahba90spline} for a detailed account), for which
efficient algorithms are available and implemented in standard software
packages. It is worth noticing that spline functions being piecewise
polynomials, once the spline has been estimated, its derivative is readily
available.

\begin{remark}
\label{rem:inv.genes}
These estimations are performed on a set of genes with small variance across
samples in each experiment, termed below {\it invariant gene set}. This gene set
must also be representative of the range of values of the average gene
expressions. Practically, the invariant gene set is determined as follows:
\begin{quote}
$\bullet$ Split the range of values of the observed gene expressions into intervals.\\
$\bullet$ Within each interval, select a given percentage of genes with smallest
variance.
\end{quote}
\end{remark}

\subsubsection{Mean and variances estimation}
\label{sub:meanvarest}
The average gene expressions $\mu_g$ are re-estimated at each step of the
algorithm as weighted sample averages of the estimated gene expressions: at
iteration $t$,
$$
\mu_g(t) = \frac1{KN_c}\sum_{k,c}x\lab(t)\ ,
$$
where $N_c=\sum_kN^k_c$ is the total number of arrays.

The variance component estimation is a difficult task here, as many gene
variances $\sigma_g^2$ are to be estimated. The I-MINQUE (or REML) approach (see
e.g.~\cite{Rao71estimation}) is a natural choice, which is unfortunately not
suitable here as it produces negative estimates for the gene variances. To
estimate the latter, we choose sample estimates from the initialization (i.e.
pre-adjusted expression data, see Section~\ref{sub:init} below).
To estimate the observation noise variances $\tau_k^2$,
we resort to the I-MINQUE approach, applied to the {\it invariant gene set}
(defined in {\sc Remark}~\ref{rem:inv.genes}).

\begin{remark}
Our simulations (see the {\em E.Coli} example below) show that when the
observation noise is large, the corresponding variances $\tau_k^2$
can be underestimated by our procedure, which in turn results in an
overestimation of the gene variances. To overcome this problem, a possible
solution is to regularize the gene estimates by adding a positive constant
$\lambda^2$ to the estimated noise variances: modified estimates of the form
$$
\hat\tau_k^2 = \tau_k^2 + \lambda^2
$$
are considered. In practical situations, since the observation noise is unknown,
the regularization parameter $\lambda$ can be used as a tuning parameter.
\end{remark}
\subsubsection{Adjusted gene expression values estimation}
\label{sub:genexpest}
Suppose now that the observation functions $f_k$ are known, the minimization of
the log posterior probability in~\eqref{eq:logposterior} leads to the
optimization of
$$
\Gamma'(\bx) = \sum_{k,g}\left[\cL^{(1);k}_g  + \cL^{(2);k}_g\right] \ ,
$$
which splits into a sum of $N_g$ decoupled terms: the estimation of the
intrinsic gene expression values $x\lab$ reduces to minimizing for each
$g=1,\dots N_g$ the quantity 
$$
\Phi_g(\bx_g)=\sum_{k,c}\left\{
\frac1{2\tau_{k}^2}\left[y\lab\! -\! f_k(x\lab)\right]^2
+\frac1{2\sigma_g^2}\left[x\lab\! -\! \mu_g\right]^2
\right\}\ ,
$$
where $\bx_g = \{x\lab,\, k=1,\dots K, c=1,\dots N_c^k\}$.

Due to the non-linearity of the observation functions $f_k$, no closed-form
expression exist for the solution, and we resort to an iterative numerical algorithm.
We assume that the mean $\mu_g$ and variances $\sigma_g^2$ and $\tau_k^2$ are
known, as well as the observation functions $f_k$. At iteration $t$, assume that
the previous estimate $x\lab(t-1)$ of the gene expression values is available. A
linearization of the observation functions $f_k$ in the neighborhood of
$x\lab(t-1)$ yields the first order approximation
$x\lab(t)= x\lab(t-1) + \epsilon\lab$, and
$$
f_k(x\lab(t)) \approx f_k(x\lab(t-1)) + \epsilon\lab f'_k(x\lab(t-1))\ ,
$$
from which we deduce
$$
\Phi_g(\bx_g(t))\approx\sum_{k,c}\bigg\{
\frac1{2\tau_{k}^2}\big[\epsilon\lab f'_k(x\lab(t\!-\!1)) - (y\lab\!
-  f_k(x\lab(t-1)))\big]^2 + \frac1{2\sigma_g^2}\big[\epsilon\lab-(\mu_g-x\lab(t-1))\big]^2
\bigg\}
$$
The update $x\lab(t)$ can therefore be obtained by optimizing with respect to
$\epsilon\lab$ the above expression,
which yields
\begin{equation*}
a\lab\epsilon\lab =
\frac{f'_k(x\lab(t\!-\!1))}{\tau_k^2}\left(y\lab\! -\! f_k(x\lab(t\!-\!1))\right)
+  \frac1{\sigma_g^2}\left(\mu_g\!-\!x\lab(t\!-\!1)\right)\ ,
\end{equation*}
where we have set\ 
$a\lab 
= [\sigma_g^2f'_k(x\lab(t-1))^2 + \tau_k^2]/\sigma_g^2\tau_k^2$.

Set now
$$
\alpha\lab = \frac1{a\lab\sigma_g^2} =
\frac1{1+f'_k(x\lab(t-1))^2\sigma_g^2/\tau_{k}^2}\ ,
$$
Then $0\le\alpha\lab\le 1$, the bounds being attained in the extreme cases (no
noise, or constant $f_k$). This yields the update rule
$x\lab(t) = x\lab(t-1) +\epsilon\lab$, i.e.
$$
x\lab(t) = \alpha\lab\mu_g + (1-\alpha\lab)\bigg[x\lab(t-1)
+ \frac1{f'_k(x\lab(t-1))}\left(y\lab\! -\! f_k(x\lab(t-1))\right)\bigg]\ .
$$
i.e. a weighted average of the mean $\mu_g$ and the contribution of
observations.
\begin{remark}
\label{rem:update.rule}
This is similar to empirical Bayes type update rules, the
difference being that the weights depend upon the observations, due to the
nonlinearity of observation functions.

The balance between the contributions of the mean and the observations is
controlled by the {\em signal to noise ratio} (SNR) $\sigma_g^2/\tau_k^2$. The
larger the SNR, the smaller the $\alpha$ parameter and the larger the
contribution of the observations: adjusted values are closer to the observed
ones. Conversely, for small SNR values, adjusted gene expressions are closer to
the means $\mu_g$.
\end{remark}
\begin{remark}
It may be shown that this update rule is actually equivalent to a variable step
gradient descent method:
$$
x\lab(t) = x\lab(t-1) + \epsilon\lab(t-1)\nabla_{x\lab}\Phi(\bx(t-1))\ ,
$$
where the stepsize depends on the gene $g$, the condition $c$, the study $k$ and
the iteration index
$$
\epsilon\lab(t-1) = \sigma_g^2\alpha\lab(t-1)\ .
$$
\end{remark}

\subsubsection{Initialization.}
\label{sub:init}
The estimation procedures described above are used within an iterative
algorithm. The latter needs either initial values for the intrinsic gene
expressions, or the observation functions, since only data $\by$ is available.

We use an estimate for the reciprocal of the observation functions (the
so-called {\em rectification functions}) provided by
the approach described in~\cite{Roubaud09approche}.
The estimation of rectification function is formulated
in~\cite{Roubaud09approche} as another smoothing spline problem:
optimize, with respect to the mean gene expressions $\mu_g$ and the
rectification functions $\varphi_k$ the quantity
$$
\sum_{k=1}^K\sum_{g=1}^{N_g}
\frac1{N_c^k}\sum_{c=1}^{N_c^k} \left[ \varphi_k(x\lab) - \mu_g\right]^2 +
\sum_{k=1}^K \lambda_k\int |\varphi_k''(x)|^2\,dx\ .
$$
The problem is solved by an iterative algorithm, in the same spirit as the
approach described here. Besides the fact that the approach
of~\cite{Roubaud09approche} estimates rectification functions instead of
observation functions, the main difference with the current approach is that the
latter models  observation noise, which is not the case of the
former. The approach developed in the present work is more complex, but is able
to correct for study-dependent observation noise.

\subsection{Algorithm}
The proposed approach can be summarized as follows:
\begin{itemize}
\item
{\bf Initialization:} Start from a first estimate $x\lab(0)$ for the intrinsic
expression values, provided by rectification function estimates
(see section~\ref{sub:init}). Estimate the gene average expressions $\mu_g$ and
variances $\sigma_g^2$ as in section~\ref{sub:meanvarest}, and the observation
functions as in section~\ref{sub:obsfuncest}.
\item
{\bf Iteration $t$:} estimates $x\lab(t-1)$ are available.
\begin{quote}
$\ast$ Re-estimate the gene expressions $x\lab(t)$ as in~\ref{sub:genexpest}.\\
$\ast$ Update the mean gene expressions $\mu_g(t)$ as in~\ref{sub:meanvarest}.
\end{quote}
\end{itemize}
The output of the algorithm consists in estimates $\hat\bx=\{\hat x\lab\}$
for the expression datasets $\bx=\{x\lab\}$, to be exploited for further
analyses, together with estimates for the means $\mu_g$, variances $\sigma_g^2$
and $\tau_k^2$, and the observation functions $f_k$.

\medskip
The algorithm was implemented using the {\tt R} statistical
environment, from which we used the smoothing spline function {\tt smooth.spline}.
Bioinformatics related functions from the {\tt Bioconductor} package were also used.

\subsection{Validation: differential analysis}
Even though differential analysis is not the only application we have in mind,
it provides a convenient setup for validating  the proposed approach. In the
numerical results we discuss below, differentially expressed genes were searched
for in original (distorted) datasets and adjusted datasets, using $t$-test
with FDR correction for multiple testing. We used functionalities from the
{\em Bioconductor} package, namely the {\tt MTP} function from the {\tt
  multtest} package.


\section{Results and discussion}
The proposed approach has been tested and validated on several problems, of
increasing complexity. We first verified that the algorithm performs well on
articifial data, i.e. random data simulated according to the above model. The
results are quite satisfactory, but since there's not much to discuss about such
simulations we refrain from reporting on them here, and prefer to focus on results
closer to practical situations. We report below on two different validations of
the approach. The first one concerns {\em semi-artificial data}, namely real
data, with clear biological outcome, which are artificially distorted according to the
distortion model above. We show that in such a situation, the proposed approach
allows one to almost completely correct for the effect of the distortion, and
recover the biological meaning of the data.
The second validation is performed on two real human gene expression datasets,
focusing on prostate cancer.

\subsection{Real data with artificial distortions}
A test was performed using artificial observation functions, applied to
real data. Namely, we chose a dataset with well understood biological
outcome, splitted it into two well balanced subsets denoted by $\bf E_{1o}$
and $\bf E_{2o}$ (``o'' stands for ``original'', see below for details) and
applied to the two subsets two different observation functions, before 
adding Gaussian observation noise (yielding distorted subsets $\bf E_{1d}$
and $\bf E_{2d}$. After adjustment, datasets denoted by $\bf E_{1a}$
and $\bf E_{2a}$ were obtained, that were aggregated.
The goal was to study the impact of the
deformation induced by the observation functions and the noise (which were
chosen so as to hide the biological effects), and the ability of the algorithm
to perform a sensible correction.

\begin{figure}
\centerline{
\includegraphics[width=7cm]{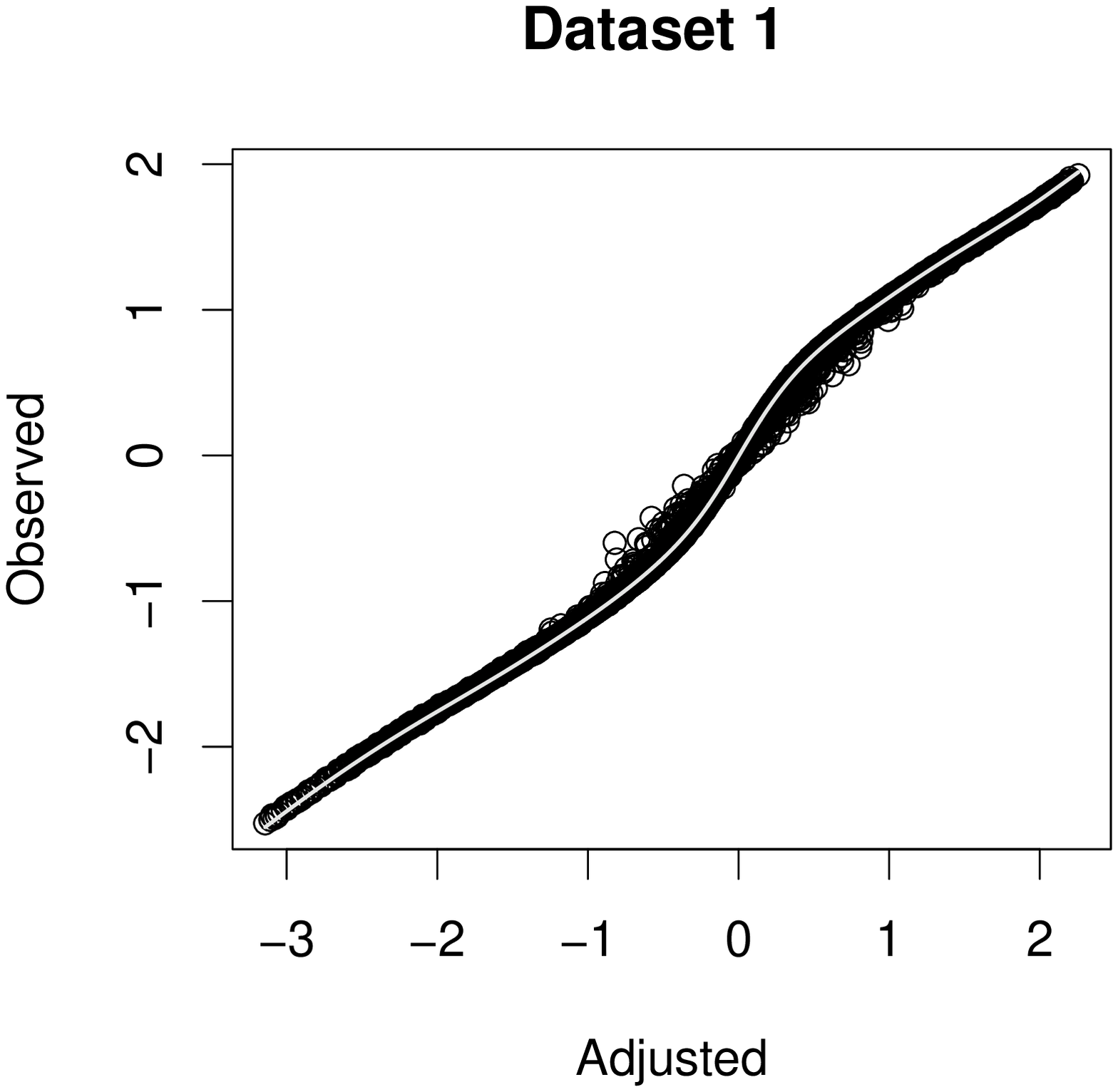}
\qquad
\includegraphics[width=7cm]{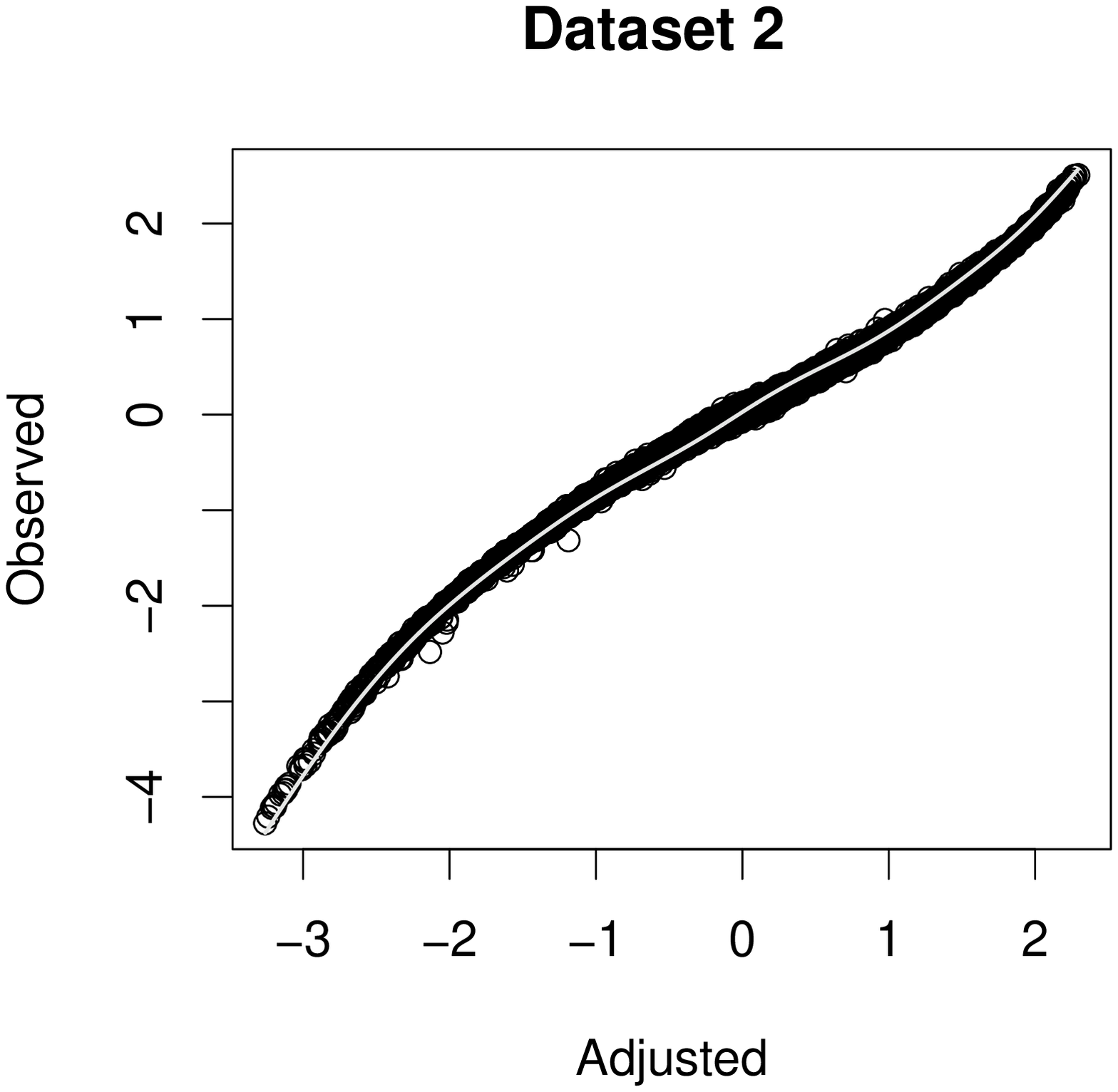}
}
\caption{Artificially distorted {\em E.Coli} dataset: observed data vs adjusted data,
  and estimated observation functions.}
\label{fi:EColi.obs.func}
\end{figure}

\begin{figure}
\centerline{
\includegraphics[width=7cm]{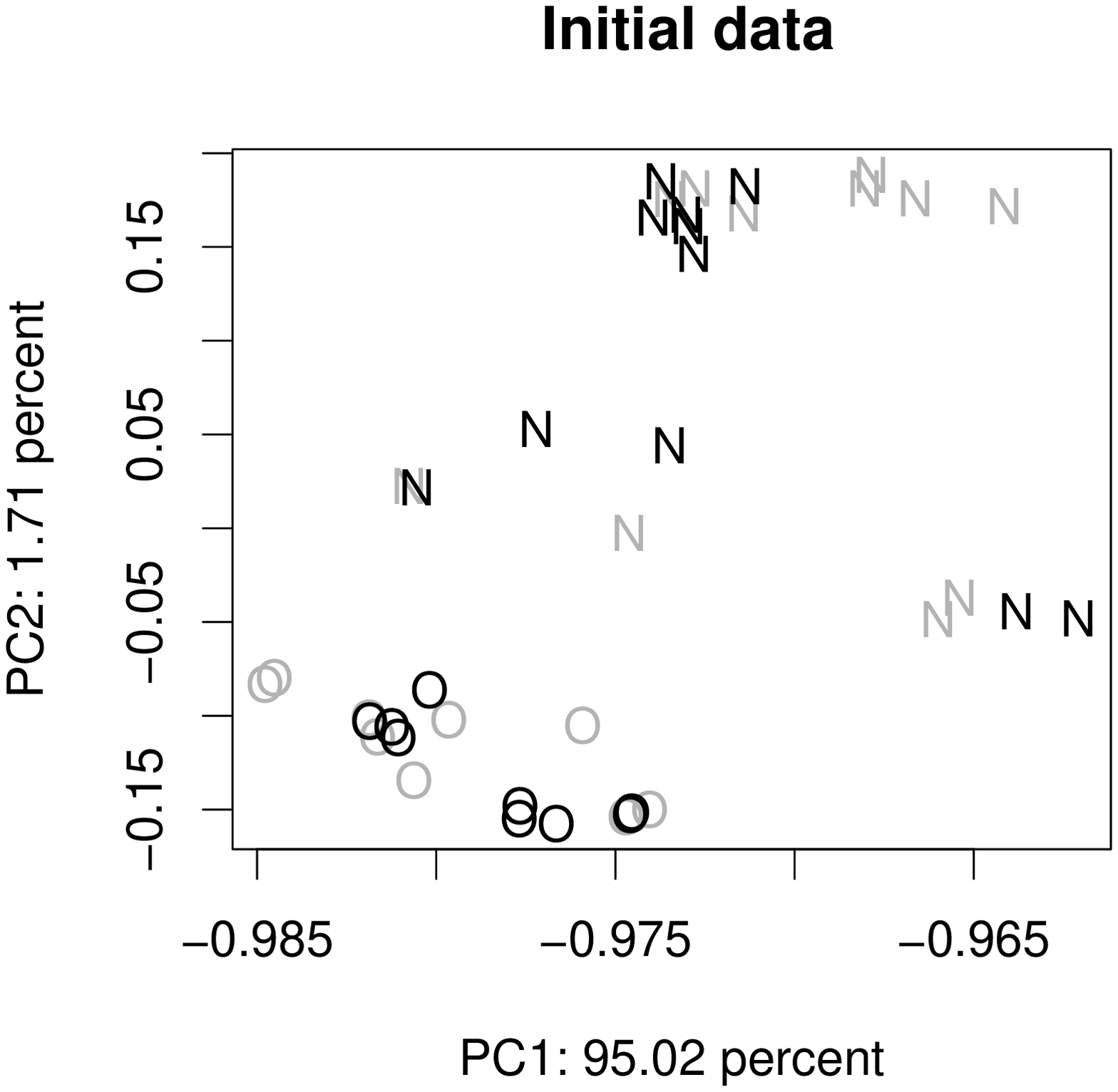}
\includegraphics[width=7cm]{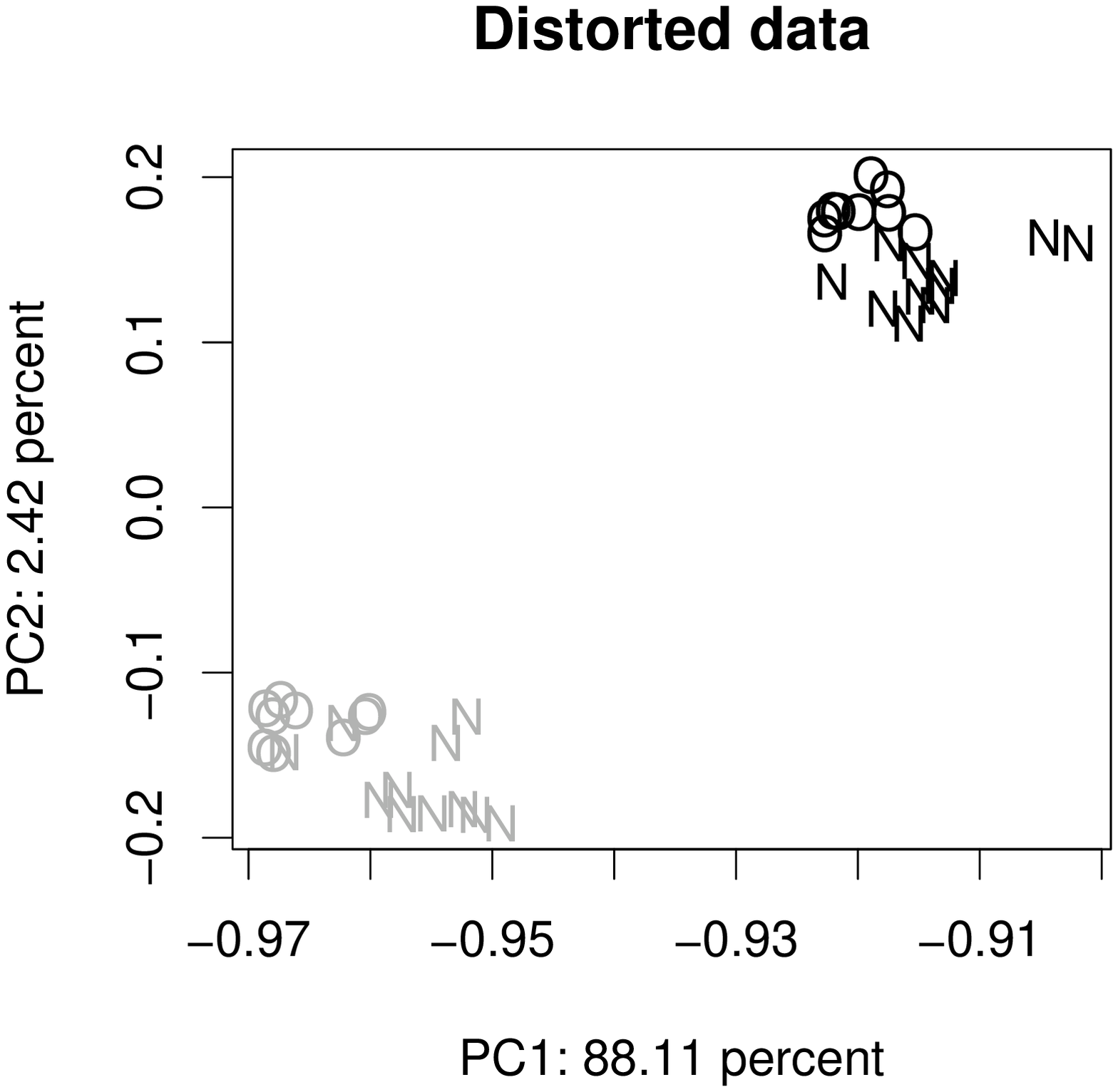}
}
\centerline{
\includegraphics[width=7cm]{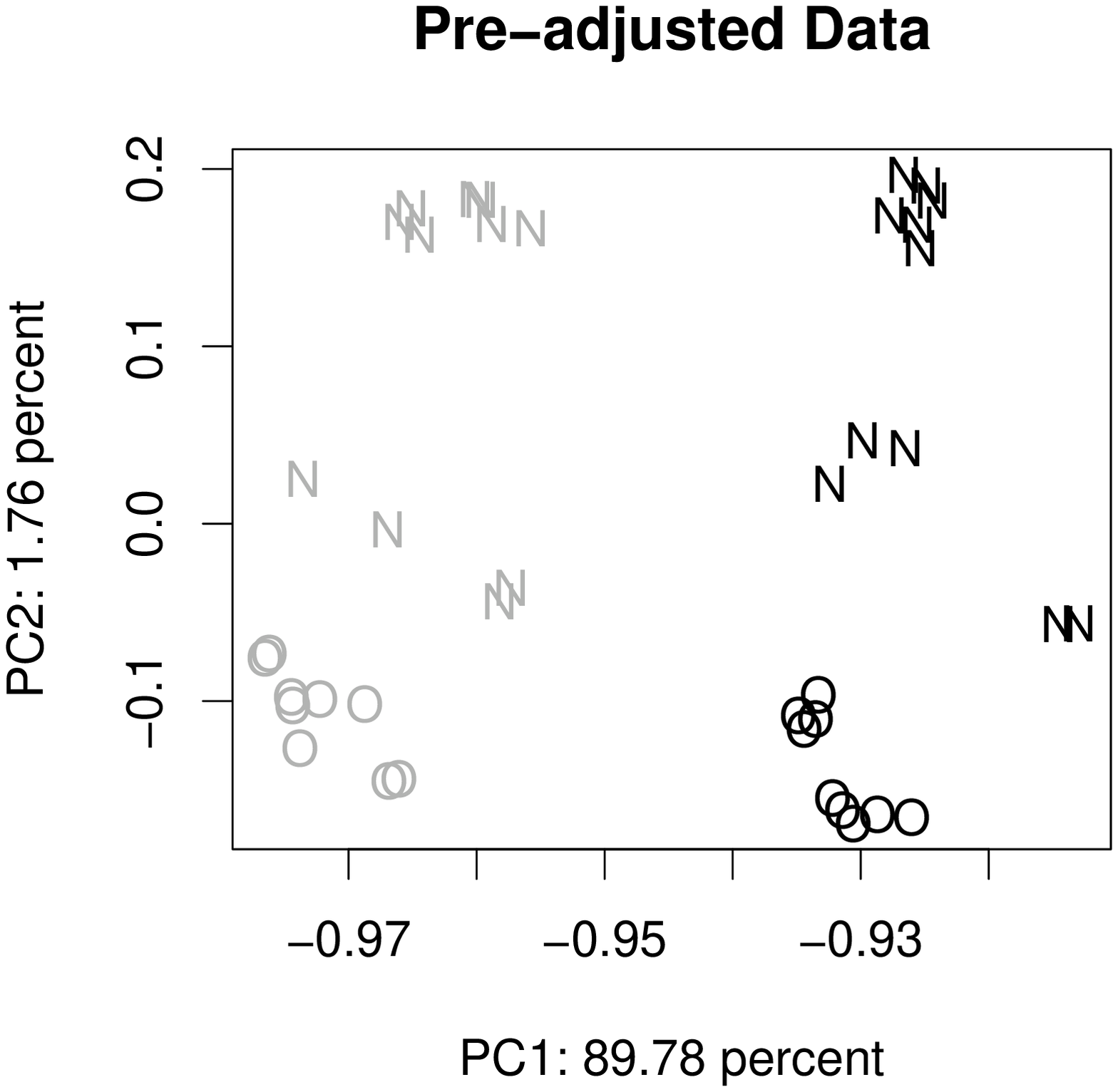}
\includegraphics[width=7cm]{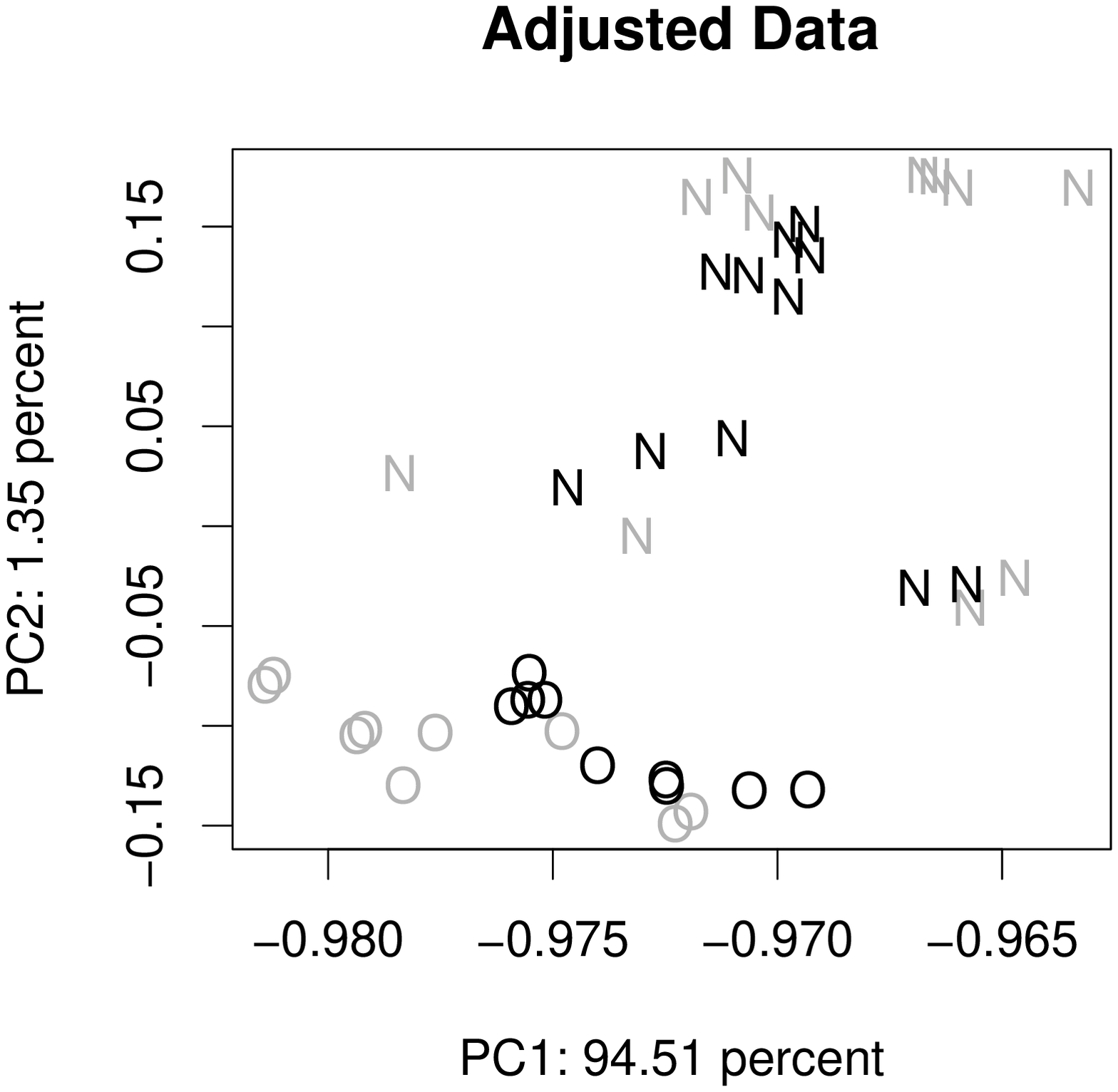}
}
\caption{Projections on the first principal plane. Top: original data (left) and
  distorted data (right). Bottom: rectified data: initialization
  (left) and processed data (right). O: aerobic; N:
  anaerobic. Gray: dataset 1; black: dataset 2}
\label{fi:SimlPCAs}
\end{figure}

{\em E. Coli} expression data from~\cite{Covert04integrating} were used. The
data include expressions of 7295 genes under two different situations (20
aerobic and 22 anaerobic). They exhibit a clear variability between the 
two biological situations, and are thus particularly interesting since they
provide a good test of the ability of our approach to recover such a variability
after distortion and correction.

Two subsets were created, both involving randomly chosen 10 aerobic and 11
anaerobic conditions. Non-linear transformations $f_1$ and $f_2$
were applied to the two so-created subsets (after standardization), and
observation noise with variances $\tau_1^2$ and $\tau_2^2$ was added. We report
here on the particular case
$$
f_1(x)=x^{0.7}\qquad\hbox{and}\qquad f_2(x)=x^{1.4}\ ,
$$
with various choices of the observation noise variances, similar results were
obtained with different choices of non-linearities.
The values for the noise variances were chosen as a function of the standard
deviations of the distorted datasets, following the rules
$$
\tau_1=\alpha\,{\rm std}(f_1(\bx_1))/10\ ,\quad \hbox{and}\quad
\tau_2=3\alpha\, {\rm std}(f_2(\bx_2))/10\ ,
$$
with various values of the tuning parameter $\alpha$. Actual
variance values, together with the corresponding estimated values, are displayed
in {\bf Table~\ref{tab:results.coli2}}, together with the estimated average gene
variance, denoted by $\langle\sigma_g^2\rangle$. As can be seen, the estimates
are fairly good, except for the last situation, which corresponds to the value
$\alpha=20$, presumably too large for our procedure (results continue to degrade
for higher values of $\alpha$, which probably originates from the degradation of
the gene variance estimates).

To illustrate graphically the effect of the adjustment,
we display in {\bf Fig.~\ref{fi:SimlPCAs}} the first
factorial plane obtained by (normalized) principal component analysis with
original data, distorted data, pre-adjusted data (using the method
of~\cite{Roubaud09approche}, used here as initialization, see
Section~\ref{sub:init}), and adjusted data.
The projection of the output data $\hat x\lab$ onto the first factorial plane
turns out to reproduce fairly accurately the projection of the original data
$x\lab$ (before distortion). The processing has therefore permitted to recover
the biological features as the first source of variability. This is confirmed by
visual inspection of the projections on other factorial planes (not shown here).

\medskip
A more quantitative validation can be obtained through differential analysis.
We searched for genes differentially expressed between aerobic and anaerobic
conditions, on initial data, distorted data and adjusted data, for various
values of observation noise. For the sake of comparison with meta-analysis based
approaches, we also performed differential analysis on the two subsets $\bf E_1$
and $\bf E_2$ (in original, distorted and adjusted situations) and computed the
intersection of the two estimated gene subsets in each situation. The goal of
the comparison is to asses the ability of each approach to recover original
differentially expressed genes.

\begin{table}[!t]
\begin{center}
{\begin{tabular}{|l||c|c|c|c|c|}
\hline
&&&&&\\
&$\tau_1^2$&$\hat\tau_1^2$&$\tau_2^2$&$\hat\tau_2^2$&$\langle\hat\sigma_g^2\rangle$\\
&&&&&\\
\hline
 S1 &$2.5\, 10^{-3}$&$3\, 10^{-3}$&$2.25\, 10^{-2}$&$2.1\, 10^{-2}$&$1.6\, 10^{-2}$\\
\hline
 S2 &$4.9\, 10^{-3}$&$5\, 10^{-3}$&$4.41\, 10^{-2}$&$4\, 10^{-2}$&$1.9\, 10^{-2}$\\
\hline
 S3 &$1\, 10^{-2}$&$7\, 10^{-3}$&$9\, 10^{-2}$&$7.1\, 10^{-2}$&$2.6\, 10^{-2}$\\
\hline
\hline
 S4 &$1.96\, 10^{-2}$&$1.3\, 10^{-2}$&$1.76\, 10^{-1}$&$1.31\, 10^{-1}$&$3.8\, 10^{-2}$\\
\hline
\end{tabular}}{}
\end{center}
\caption{\small Observation noise and average gene variance for the four simulated datasets and their estimation.}
\label{tab:results.coli2}
\end{table}

\begin{table}[!t]
\begin{center}
{\begin{tabular}{|l||c||c|c||c|c|c|}
\hline
&&&&&&\\
&\bf O&\bf D&\bf A&\!\!\bf E\sbs{1o}$\cap$E\sbs{2o}\!\!&\bf
\!\!E\sbs{1d}$\cap$E\sbs{2d}\!\!&\!\!\bf E\sbs{1a}$\cap$E\sbs{2a}\!\!\\
&&&&&&\\
\hline
\bf S1+&258&121 (129)&162 (173)&5&1&1\\
\bf S1-&328&139 (141)&187 (188)&31&9&10\\
\hline
\bf S2+&275&107 (112)&123 (129)&5&2&2\\
\bf S2-&312&124 (129)&148 (155)&31&3&4\\
\hline
\bf S3+&238&71 (74)&88 (93)&5&1&1\\
\bf S3-&305&105 (109)&113 (116)&30&3&4\\
\hline
\hline
\bf S4+&253&58 (60)&65 (68)&7&1&1\\
\bf S4-&300&72 (73)&80 (81)&32&1&1\\
\hline
\end{tabular}}{}
\end{center}
\caption{\small Differential analysis on artificially distorted {\em E.Coli} dataset.
The first column indicates the simulation index (see
Table~\protect{\ref{tab:results.coli2}} for the corresponding observation noise
variances), the sign indicates over or under expression in aerobic conditions.
Column 2: numbers of differentially expressed genes in the original dataset.
Columns 3 to 7: numbers of differentially expressed genes in distorted
({\bf D}) and adjusted ({\bf A}) datasets, and in the intersections
$\bf E_{1o}\cap E_{2o}$, $\bf E_{1d}\cap E_{2d}$ and $\bf E_{1a}\cap E_{2a}$
that are present in the original dataset (the numbers between parentheses
indicate the total numbers of differential genes).
\label{tab:results.coli}}
\end{table}

The results are displayed in {\bf Table~\ref{tab:results.coli}}, which gives
the numbers of differentially expressed genes on the original dataset (first
column), and the number of original differential genes that are recovered
on the distorted dataset (second column) and the adjusted dataset
(third column); total differential gene numbers are given between parenthesis.

A first conclusion is that the distortion significantly degrades the
differential analysis, the degradation being obviously bigger when the
observation noise is higher. The processing improves the situation, which is
reflected by an average increase of the number of the recovered
original differential genes ranging between 11\% and 20\% (depending on the
noise level). 

\medskip
For the sake of comparison, we also display in the last three columns numbers of
differential genes obtained by intersecting the results of individual
differential analysis on the two sub-datasets, in the three considered
situations. It is worth noticing that the loss of differential genes is in this
case really huge, which is for us a strong motivation for aggregating datasets
rather than intersecting differential analysis results.

\subsection{Real data}

The proposed approach was also tested on the two publicly available datasets of
prostate cancer expression data reported in~\cite{Singh02gene}
and~\cite{Stuart04silico}. Both datasets originate from Affymetrix HG-U95av2
arrays.

We started from raw expression data, and normalized them separately using the
standard RMA normalization procedure from the Bioconductor package.
After normalization, the dataset of~\cite{Singh02gene}
appeared to exhibit much larger variability than
the~\cite{Stuart04silico} dataset. We thus performed the following
pre-processing: we extracted a subset of arrays whose correlations
to the median array exceed 90\%, and reduction to common genes).
After pre-processing, the two-datasets consist in respectively 61 (32 tumor and
29 normal) and 86 (37 tumor and 49 normal) conditions, with 12625 genes, from
which we extracted balanced sub-datasets (29 tumor and 29 normal for each)

\begin{figure}[h]
\centerline{
\includegraphics[width=7cm]{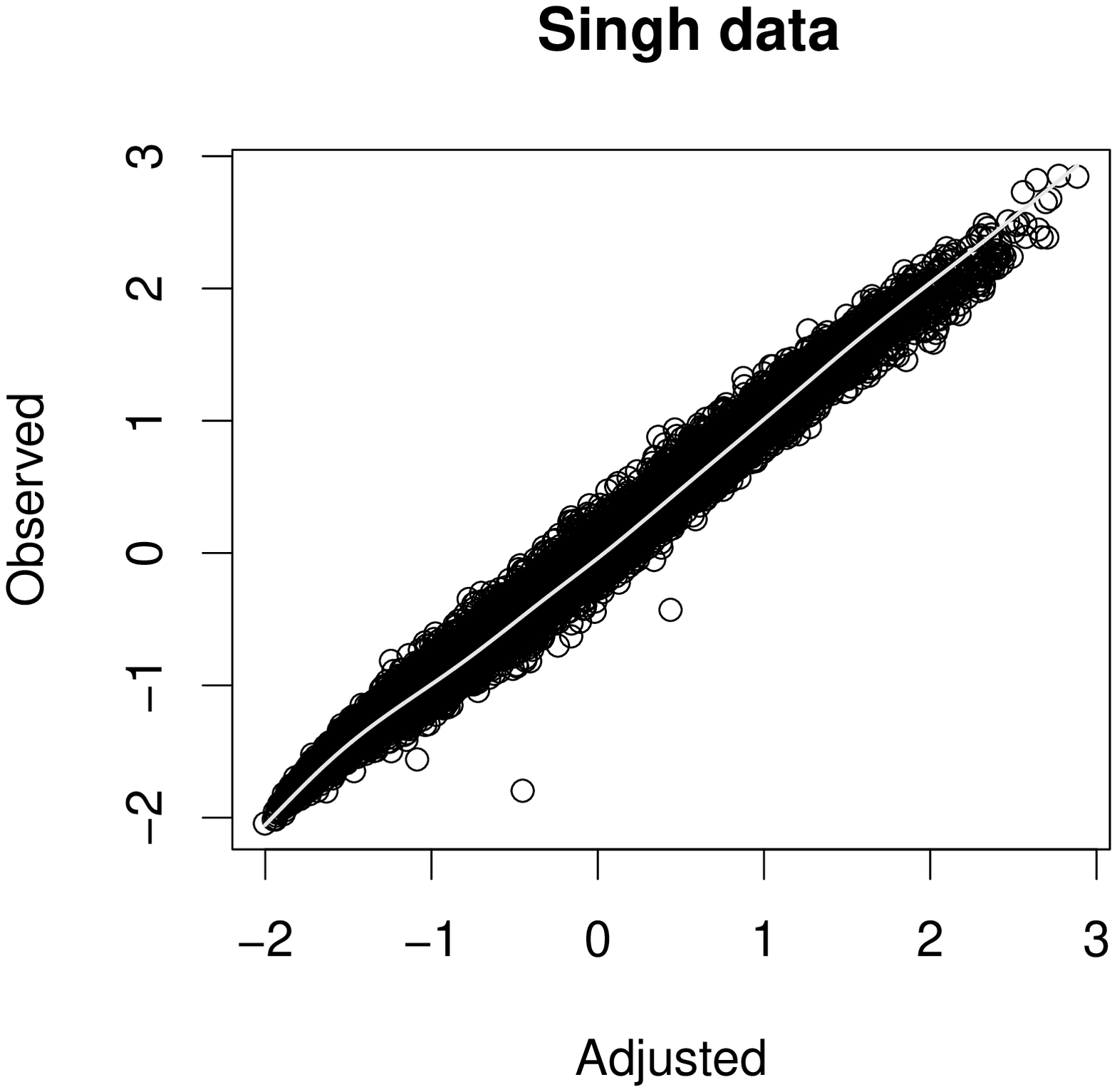}
\quad
\includegraphics[width=7cm]{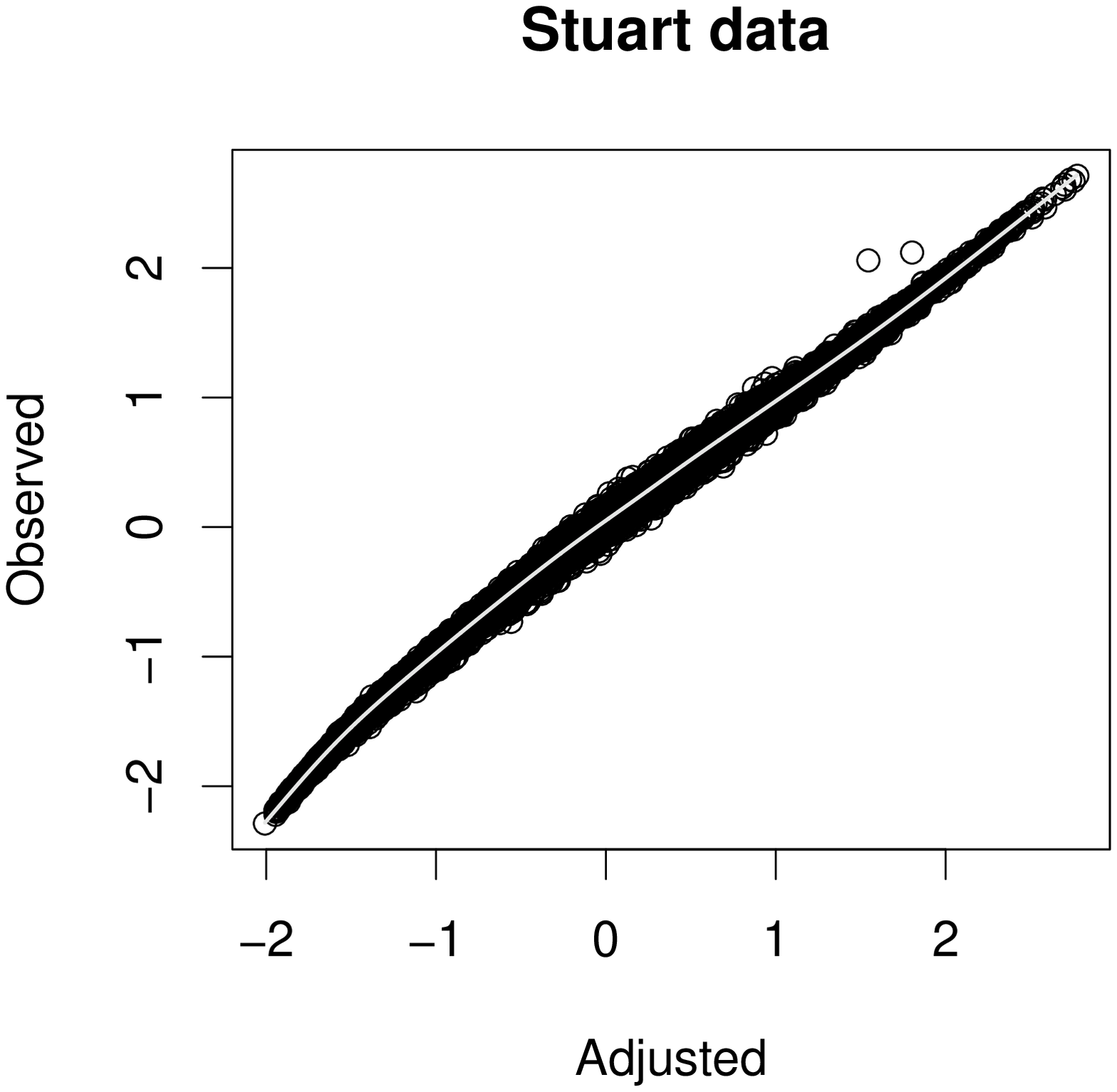}
}
\caption{Estimated observation functions for the prostate datasets: Singh (left)
  and Stuart (right) datasets}
\label{fi:prostate.obsfun}
\end{figure}

\begin{figure}
\centerline{
\includegraphics[width=7cm]{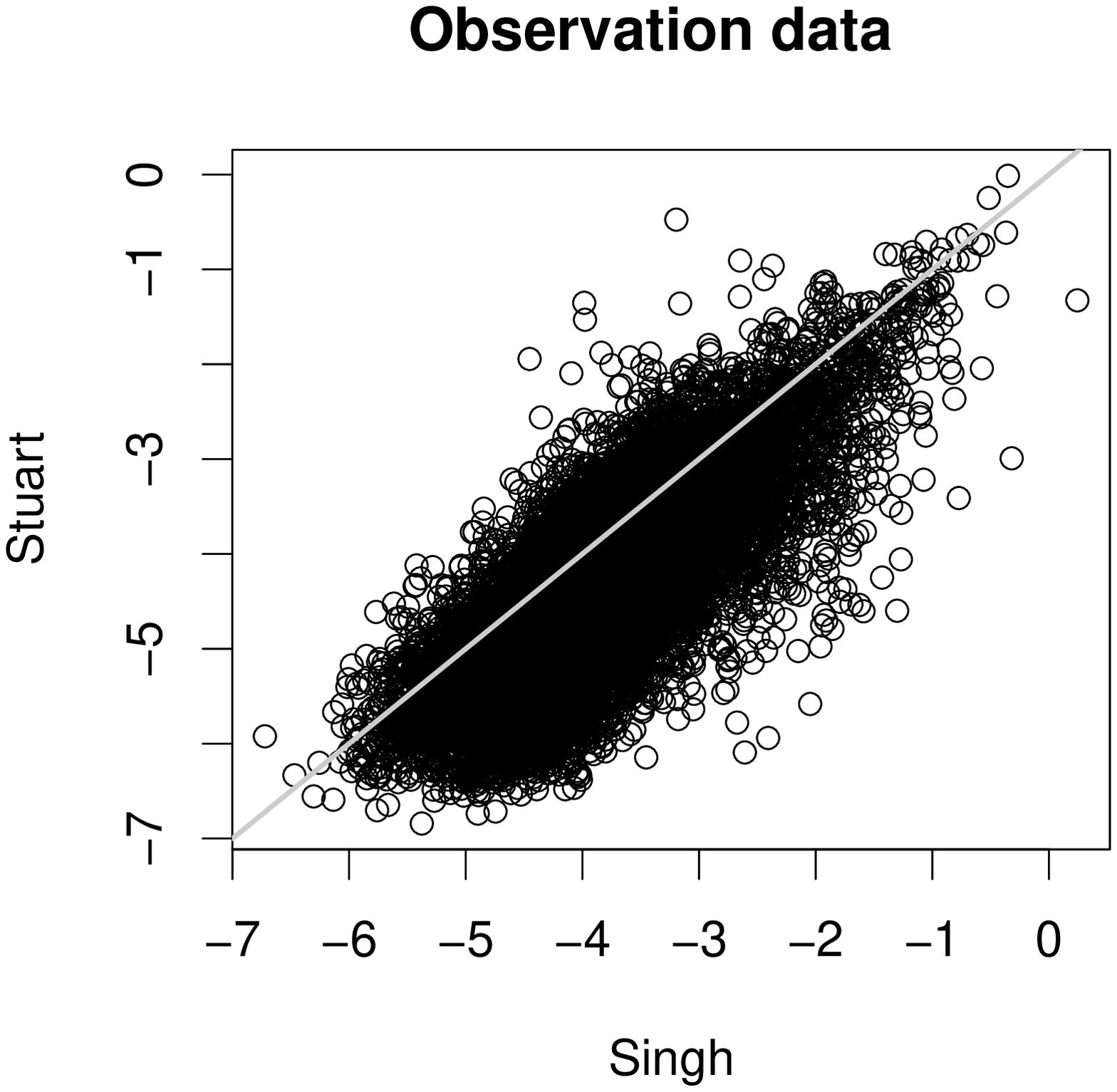}
\quad
\includegraphics[width=7cm]{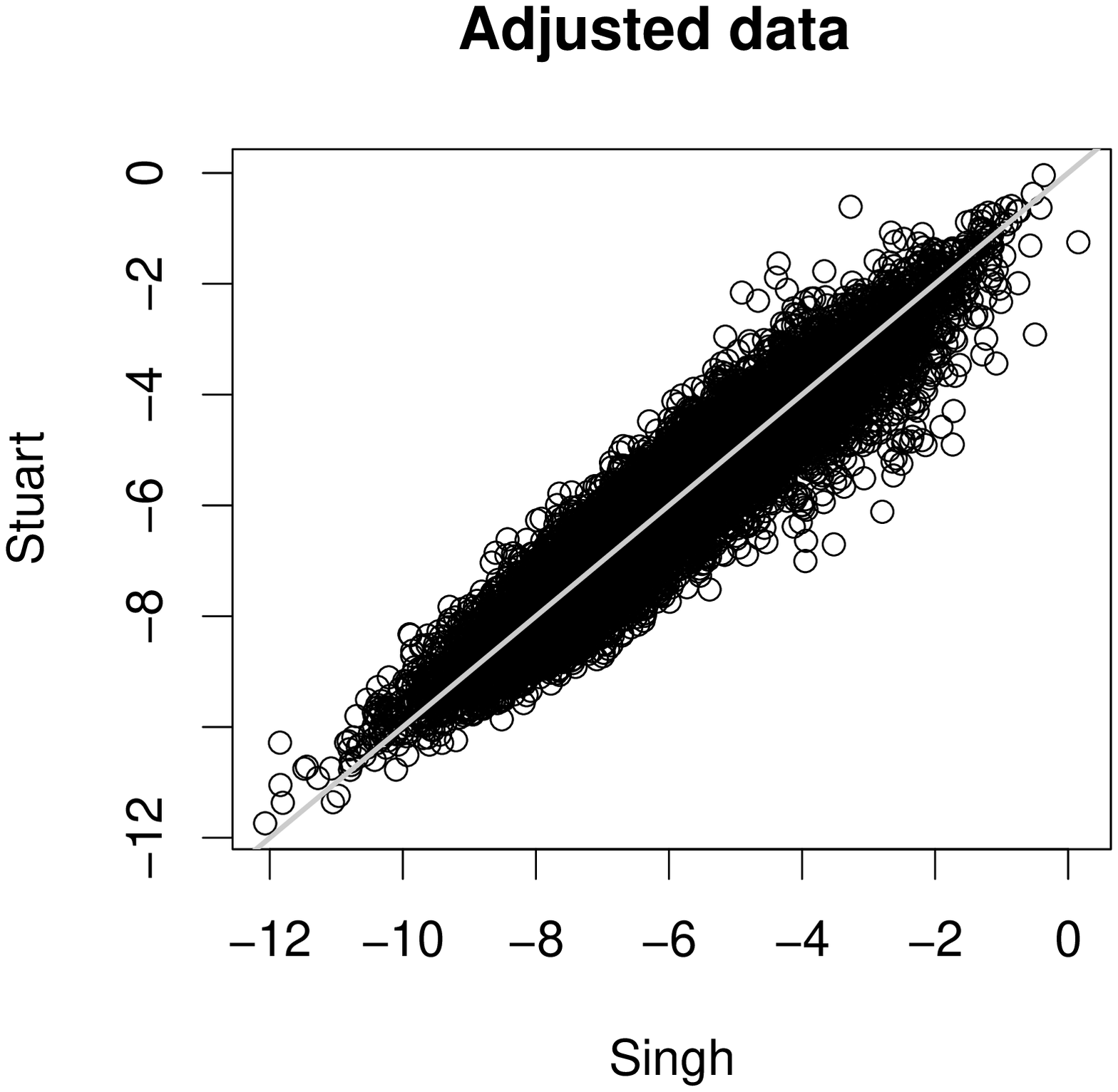}
}
\caption{Prostate datasets: estimated gene variances in the original dataset
  (left) and the adjusted dataset (right), in logarithmic scales}
\label{fi:prostate.variances}
\end{figure}

The proposed algorithm was run on these two datasets. The result
of the processing is shown in Fig.~\ref{fi:prostate.obsfun} (estimates for the
observation functions) and in Fig.~\ref{fi:prostate.variances} (estimated
variances, for the observed and adjusted data). We first notice that the
estimated observation functions are in this case fairly regular, no large
distortion has been found.  It can be seen on Fig.~\ref{fi:prostate.obsfun}
that the Singh data (termed {\textbf E\sbs{1}}) have been shrunk more
importantly than the Stuart data  ({\textbf E\sbs{2}}). 
This can be interpreted in terms of the observation noise. Indeed, the
observation noise was found to be significantly larger in the Singh dataset
than in the Stuart dataset:
$\hat\tau_1^2\approx 2.6\,10^{-2}$ and $\hat\tau_2^2\approx 1.6\,10^{-2}$.
Nevertheless, both values are larger than the estimated average gene variance
$\langle\hat\sigma^2\rangle\approx 4.2\,10^{-3}$.
According to Remark~\ref{rem:update.rule}, we are here in a situation where the
contribution of the average to the adjusted gene expression values tends to be
larger than the contribution of the observations.

Finally, it is clear from
Fig.~\ref{fi:prostate.variances} that the proposed approach has done quite a
good job for adjusting the gene variances: variances in adjusted Singh and
Stuart datasets are coherent. The interstudy variability has
therefore been considerably reduced.

\medskip
As in the previous case study, to assess more quantitatively the performances of
the proposed approach, differential analysis was performed on the real dataset,
and the adjusted dataset. 
After filtering out the 30\% least variable genes,
we used $t$-test based decision rule, with FDR correction for
multiple testing ($\alpha=5\%$, 5000 bootstrap samples).
Differentially expressed genes were seeked for the real dataset and the
processed dataset, as well as the individual subsets {\bf E\sbs{1}}
(Singh) and {\bf E\sbs{2}} (Stuart),  and the corresponding adjusted
subsets {\bf E\sbs{1a}} and {\bf E\sbs{2a}}. The results are summarized
in {\bf Table~\ref{tab:results.prostate}}. 

\begin{table}[!h]
\begin{center}
{\begin{tabular}{|l||c||c||c|c|c||c|c|}
\hline
&&&&&&&\\
&\bf O&\bf A&\bf O$\cap$\bf A&\bf O$\backslash$ A&\bf A$\backslash$ O
&\!\!\bf E\sbs{1}$\cap$E\sbs{2}\!\!&\!\!\bf E\sbs{1a}$\cap$E\sbs{2a}\!\!\\
\hline
\bf +&73&88&66&7&22&4&5\\
\hline
\bf -&232&245&217&15&28&1&1\\
\hline
\end{tabular}}{}
\end{center}
\caption{\small Differential analysis on the {\em Prostate} dataset: numbers of
  differentially expressed genes. First column: aggregation of the two original
  datasets. Second column: aggregation of the two adjusted datasets. Columns
  from 3 to 5: intersection and differences between original and adjusted.
  Columns 6 and 7: intersections of original subdatasets and adjusted
  subdatasets respectively.
\label{tab:results.prostate}}
\end{table}

\begin{figure}
\centerline{
\includegraphics[width=10cm]{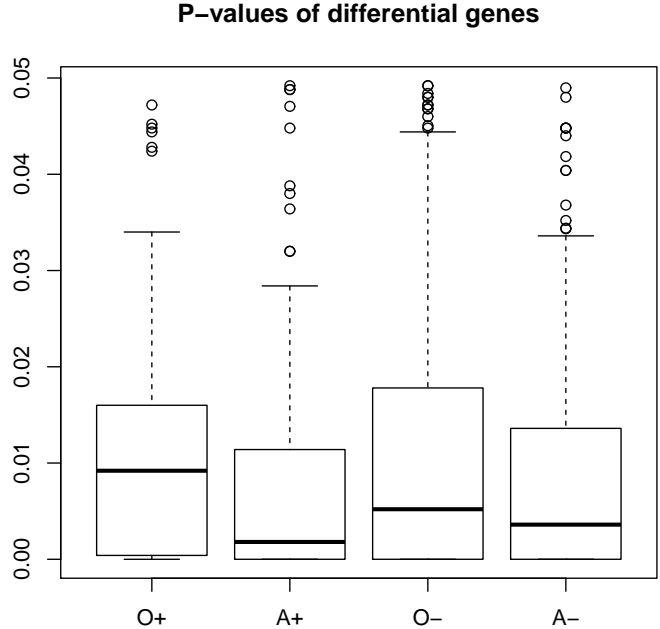}
}
\caption{P-values of differential genes for the original and the adjusted
  datasets: Positively differential genes in the original (O+) and adjusted
  (A+) datasets, and negativeley differential genes in the original (O-) and
  adjusted (A-) datasets}
\label{fi:DiffGenesPvalues}
\end{figure}

As before, a striking first result is the fact that the intersection of the
differential genes found from {\bf E\sbs{1}} and {\bf E\sbs{2}} individually is
extremely small, as well as the the intersection of the differential genes
found from {\bf E\sbs{1a}} and {\bf E\sbs{2a}} individually. Quite a large
number of differential genes appear to be lost by this procedure.
Besides, as was also seen in the above {\em E. Coli} simulation study, the
adjusted datasets yield a larger number of differential genes than the original
datasets. A closer examination of the results of the differential analysis leads
to the following conclusions:
\begin{itemize}
\item
The adjusted dataset features, after concatenation, a larger number of
positively and negatively differential genes. Even if one focuses on
differential genes with small p-values (say, less than 1\%), the adjusted
dataset yields a significant number of ``new'' genes, including for example
MAPKAPK3, POLR2H, BDH1, PYCR1, PDIA4, SLC25A6, ZMPSTE21, ENTPD6 for the
positively differential genes (i.e. overexpressed in tumors), and ADD1, VAT1,
PPP2CB, RBPMS, FSCN1, ARHGEF4, GRK5, FEZ1, HTRA1, RBMS1, CETN2 and OSR2 for the
negatively differential genes.

A precise analysis of the role of these genes in prostate cancer goes far beyond
the scope of the present work. Let us simply notice that most of these genes
have already been reported in the prostate cancer literature. This makes them
potentially interesting.
\item
Globally, the p-values of differential genes are smaller in the adjusted dataset
than in the original dataset. This appears clearly in {\bf
  Fig.}~\ref{fi:DiffGenesPvalues} where they are displayed in boxplot forms.
\item
Conversely, less genes were found differential in the original dataset only; in
addition, the corresponding p-values are generally close to the critical value
chosen in this study (5\%). A closer examination shows that their variance in
the adjusted dataset is smaller than in the original one. According to
Remark~\ref{rem:update.rule}, this presumably originates from a low signal to
noise ratio $\sigma_g^2/\tau_k^2$, or a large value of the observation function
derivative $f'(\mu_g)$ (in the present approach, these quantities govern the
parameter $\alpha\lab$ which controls the variance in the adjusted dataset).
\end{itemize}

\section{Conclusion}
We have proposed in this paper a new approach for aggregating gene expression
datasets originating from different studies. The rationale of the method is to
adjust the datasets prior to merging, using a Bayesian modeling.
The model we develop is specifically adapted to microarray data, and assumes that
the inter-study variability originates from non linear distortions, combined
with study dependent observation noise. The inter-study variability model is
combined with a Gaussian model for the gene expression values (after logarithmic
transformation) in a Bayesian framework.

Our results on artificial and real data show that the method is sound, and
capable of correcting for study-dependent distortion and observation noise. For
the sake of simplicity, we focused on the case of two datasets to be merged,
however the model is versatile enough to accomodate arbitrary numbers of
datasets. The results on real data discussed here are quite encouraging, however
the approach will have to be tested further on a larger scale to be validated
more thoroughly.

Several aspects are to be investigated further. Among them, the problem of
variance components estimation is an important one, as the estimated gene
variances have a key impact on the quality of the results. For the sake of
simplicity we have limited ourselves here to a simple procedure, involving
sample estimates of gene variances from pre-adjusted data. More sophisticated
approaches will have to be investigated.



\bibliographystyle{plain}
\bibliography{MCB}

\end{document}